# MUSIC OF BRAIN AND MUSIC ON BRAIN: A NOVEL EEG SONIFICATION APPROACH

Sayan Nag[1], Shankha Sanyal[2,3*], Archi Banerjee[2,3], Ranjan Sengupta[2] and Dipak Ghosh[2]

[1] Department of Electrical Engineering, Jadavpur University

[2] Sir C.V. Raman Centre for Physics and Music, Jadavpur University

[3] Department of Physics, Jadavpur University

* Corresponding Author

**ABSTRACT**

*Can we hear the sound of our brain? Is there any technique which can enable us to hear the neuro-electrical impulses originating from the different lobes of brain? The answer to all these questions is YES. In this paper we present a novel method with which we can sonify the Electroencephalogram (EEG) data recorded in rest state as well as under the influence of a simplest acoustical stimuli - a tanpura drone. The tanpura drone has a very simple yet very complex acoustic features, which is generally used for creation of an ambiance during a musical performance. Hence, for this pilot project we chose to study the correlation between a simple acoustic stimuli (tanpura drone) and sonified EEG data. Till date, there have been no study which deals with the direct correlation between a bio-signal and its acoustic counterpart and how that correlation varies under the influence of different types of stimuli. This is the first of its kind study which bridges this gap and looks for a direct correlation between music signal and EEG data using a robust mathematical microscope called Multifractal Detrended Cross Correlation Analysis (MFDXA). For this, we took EEG data of 10 participants in 2 min 'rest state' (i.e. with white noise) and in 2 min 'tanpura drone' (musical stimulus) listening condition. Next, the EEG signals from different electrodes were sonified and MFDXA technique was used to assess the degree of correlation (or the cross correlation coefficient $γ_x$) between tanpura signal and EEG signals. The variation of $γ_x$ for different lobes during the course of the experiment also provides major interesting new information. Only music stimuli has the ability to engage several areas of the brain significantly unlike other stimuli (which engages specific domains only).*

**INTRODUCTION**

Can we hear our brain? If we can, how will it sound? Will the sound of our brain be different from one cognitive state to another? These are the questions which opened the vistas of a plethora of research done by neuroscientists to sonifiy obtained EEG data. The first thing to be addressed while dealing with sonification is what is meant by sonification. As per the definition of ICAD "the use of non-speech audio to convey information; more specifically sonification is the transformation of data relations into perceived relations in an acoustic signal for the purposes of facilitating communication or interpretation" [1]. Hermann [2] gives a more classic definition for sonification as "a technique that uses data as input, and generates sound signals (eventually in response to optional additional excitation or triggering)".

The idea of making electroencephalographic (EEG) signals audible accompanied brain imaging development from the very first steps in the early 1930s. Prof. Edgar Adrian listened to his own EEG signal while replicating Hans Bergers experiments [3]. Real time EEG sonification enjoys a wide number of applications including diagnostic purposes like epileptic seizure detection, different sleep states etc. [4-6], neuro-feedback applications [7,8], brain-controlled musical instruments [9] while a special case involves converting brain signals directly into meaningful musical compositions [10,11]. Also, we have a number of studies which deal with the emotional appraisal in human brain corresponding to a wide variety of musical clips using EEG/fMRI techniques [12-14], but none of them provides a direct correlation between the music sample used and the EEG signal generated using the music as a stimulus. although both are essentially complex time series variations. The main reason behind this lacunae is the disparity between the sampling frequency of music signals and EEG signals (which are of much lower sampling frequency). EEG signals are lobe specific and characterized with a lot of variations corresponding to different musical and other stimulus. So the information procured varies continuously throughout a period of data acquisition and that too the fluctuations are different in different lobes.

In this work, the main attempt is to device a new methodology which looks to obtain a direct correlation between the external musical stimuli and the corresponding internal brain response using latest state of the art non-linear tools for characterization of bio-sensor data. For this, we chose to study the EEG response corresponding to the simplest (and yet very complex) musical stimuli - the *Tanpura* drone. The *Tanpura* (sometimes also spelled Tampura or Tambura) is an integral part of classical music in India. It is a fretless musical instrument. It consists of a large gourd and a long voluminous wooden neck which act as resonance bodies with four or five metal strings supported at the lower end by a meticulously curved bridge made of bone or ivory. The strings are plucked one after the other in cycles of few seconds generating a buzzing drone sound. The Tanpura drone primarily establishes the "Sa" or the scale in which the musical piece is going to be sung/played. One complete cycle of the drone sound usually comprises of Pa/Ma (middle octave) — Sa (upper octave) — Sa (upper octave) — Sa (middle octave) played in that order. The drone signal has repetitive quasi-stable geometric forms characterized by varying complexity with prominent undulations of intensity of different harmonics. Thus, it will be quite interesting to study the response of brain simultaneously to a simple drone sound using different non-linear techniques. This work is essentially a continuation of our work using MFDFA technique on drone-induced EEG signals [15]. Because there is a felt resonance in perception, psycho-acoustics of Tanpura drone may provide a unique window into the human psyche and cognition of musicality in human brain.

In this work, we took EEG data of 10 naive participants while they listened to the 2 min *tanpura* drone clip which was preceded by a 2 min resting period. The main constraint in establishing a direct correlation between EEG signals and the stimulus sound signal is the disparity in sampling frequency of the two; while an EEG signal is generally sampled at up to 512 samples/sec (in our case it is 256 samples/sec), the sampling frequency of a normal recorded audio signal is 44100 samples/sec. Hence the need arises to upsample the EEG signal to match the sampling frequency of an audio signal so that the correlation between the two can be established. This phenomenon is called sonification in essence and we propose a novel algorithm in this work to sonify EEG signals and then to compare them with the source sound signals. We used a robust non-linear technique called Multi Fractal Detrended Cross Correlation Analysis (MFDXA) [16] in this case, taking the *tanpura* drone signal as the first input and a music induced modulated EEG signal (electrode wise) as the second input. The output is $\gamma_x$ (or the cross-correlation coefficient) which determines the degree of cross-correlation of the two signals taken. For the "no music/rest" state, we have determined the cross-correlation coefficient using the "rest" EEG data as one input and a simulated "white noise" as the other input. We have provided a comparative analysis of the variation of correlation between the "rest" state EEG and the "music induced" EEG signals. Furthermore, the degree of cross-correlation between different lobes of the brain have also been computed for the two experimental conditions. The results clearly indicate a significant rise in the correlation during the music induced state compared to the rest state. This novel study can have far reaching conclusion in the domain of auditory neuroscience.

**MATERIALS AND METHODS:**
**Subjects Summary**
10 young musically untrained right handed adults (6 male and 4 female) voluntarily participated in this study. Their ages were between 19 to 25 years (SD=2.21 years). None of the participants reported any history of neurological or psychiatric diseases, nor were they receiving any psychiatric medicines or using a hearing aid. Informed consent was obtained from each subject according to the ethical guidelines of the Ethical Committee of Jadavpur University. All experiments were performed at the Sir C.V. Raman Centre for Physics and Music, Jadavpur University, Kolkata.

**Experimental Details**
The tanpura stimuli given for our experiment was the sound generated using software 'Your Tanpura' in C# pitch and in Pa (middle octave) — Sa (middle octave) — Sa (middle octave) — Sa (lower octave) cycle/format. From the complete recorded signal a segment of about 2 minutes was cut out at the zero point crossing using open source software toolbox Wavesurfer [35]. Variations in the timbre were avoided as same signal were given to all the participants. **Fig. 1** depicts a 2 min Tanpura drone signal that was given as an input stimulus to all the informants.

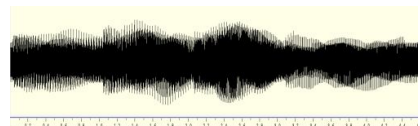

**Fig.1:** Waveform of 1 cycle of tanpura drone signal

**Experimental Protocol**

The EEG experiments were conducted in the afternoon (around 2 PM) in an air conditioned room with the subjects sitting in a comfortable chair in a normal diet condition. All experiments were performed as per the guidelines of the Institutional Ethics Committee of Jadavpur University. All the subjects were prepared with an EEG recording cap with 19 electrodes (Fig.2) (Ag/AgCl sintered ring electrodes) placed in the international 10/20 system. Impedances were checked below 5 kOhms. The EEG recording system (Recorders and Medicare Systems) was operated at 256 samples/s recording on customized software of RMS. The data was band-pass-filtered between 0.5 and 70 Hz to remove DC drifts and suppress the 50Hz power line interference. After initialization, a 6 min recording period was started, and the following protocol was followed: 1. 2 min Rest (No music) => 2. 2 min With *tanpura* drone => 3. 2 min Rest (After Music). We divided each of the experimental conditions in four windows of 30 seconds each and calculated the cross-correlation coefficient for each window corresponding to the frontal electrodes.

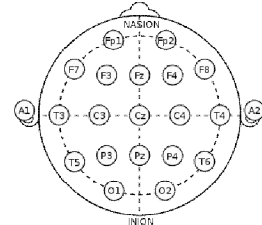

**Fig. 2** The position of electrodes as per 10-20 system

# METHOD OF ANALYSIS

**Sonification of EEG signals:**

The sampling rate of the acquired EEG signal is 256 Hz (as per the data we have used for this purpose) while the music signals used in our study is 44.1 kHz and we envisage to get a direct correlation between these two signals. No such direct correlation has been established between the cause and the effect because the EEG signals have a frequency much less than that of the music signals. Keeping in mind this problem, we up-sampled the EEG data to 44.1 kHz. The up-sampling changed an EEG signal of frequency 256 Hz to a modulated EEG signal of frequency 44.1 kHz. Noise introduced in the data are removed by filtering of the data. We used a band-pass filter for this purpose. We have added aesthetic sense to these modulated signals by assigning different frequencies or tones (electrode wise) to them so that when they are played, say for Frontal (F4) electrode the sudden increase or decrease in the data information can be perceived as manifested by the change in the amplitudes of the signal. Now that we have a modulated music induced EEG signal of the same frequency as that of the music signal we can now use cross-correlation techniques to establish any relation between these signals. We used a robust non-linear technique called Multi Fractal Detrended Cross Correlation Analysis (MFDXA) [16] in this case, taking the *tanpura* drone signal as the first input and a music induced modulated EEG signal (electrode wise) as the second input.

**Multifractal Detrended Cross Correlation Analysis (MF-DXA):**

We have performed a cross-correlation analysis of correlation between the *tanpura* drone signals and EEG signals following the prescription of Zhou [16]. Also the EEG signals from different lobes of the brain were also analyzed using the same technique.

$$x_{avg} = 1/N \sum_{i=1}^{N} x(i) \text{ and } y_{avg} = 1/N \sum_{i=1}^{N} y(i) \qquad (1)$$

Then we compute the profiles of the underlying data series x(i) and y(i) as

$$X(i) \equiv [\sum_{k=1}^{i} x(k) - x_{avg}] \text{ for } i = 1 \ldots N \qquad (2)$$

$$Y(i) \equiv [\sum_{k=1}^{i} x(k) - x_{avg}] \text{ for } i = 1 \ldots N \qquad (3)$$

The qth order detrended covariance Fq(s) is obtained after averaging over 2Ns bins.

$$F_q(s) = \{1/2N_s \sum_{v=1}^{2Ns} [F(s,v)]^{q/2}\}^{1/q} \qquad (4)$$

where q is an index which can take all possible values except zero because in that case the factor 1/q blows up. The procedure can be repeated by varying the value of s. Fq(s) increases with increase in value of s. If the series is long range power correlated, then Fq(s) will show power law behavior

$$F_q(s) \sim s^{\lambda(q)}.$$

Zhou found that for two time series constructed by binomial measure from p-model, there exists the following relationship [16]:

$$\lambda(q=2) \approx [h_x(q=2) + h_y(q=2)]/2. \tag{5}$$

Podobnik and Stanley have studied this relation when q = 2 for monofractal Autoregressive Fractional Moving Average (ARFIMA) signals and EEG time series [17].

In case of two time series generated by using two uncoupled ARFIMA processes, each of both is autocorrelated, but there is no power-law cross correlation with a specific exponent [53]. According to auto-correlation function given by:

$$C(\tau) = \langle [x(i+\tau) - \langle x \rangle][x(i) - \langle x \rangle] \rangle \sim \tau^{-\gamma}. \tag{6}$$

The cross-correlation function can be written as

$$C_x(\tau) = \langle [x(i+\tau) - \langle x \rangle][y(i) - \langle y \rangle] \rangle \sim \tau^{-\gamma_x} \tag{7}$$

where $\gamma$ and $\gamma_x$ are the auto-correlation and cross-correlation exponents, respectively. Due to the non-stationarities and trends superimposed on the collected data direct calculation of these exponents are usually not recommended rather the reliable method to calculate auto-correlation exponent is the DFA method, namely $\gamma = 2 - 2h(q=2)$ [18]. Recently, Podobnik et al., have demonstrated the relation between cross-correlation exponent, $\gamma_x$ and scaling exponent $\lambda(q)$ derived from $\gamma_x = 2 - 2\lambda(q=2)$ [17]. For uncorrelated data, $\gamma_x$ has a value 1 and the lower the value of $\gamma$ and $\gamma_x$ more correlated is the data. In general, $\lambda(q)$ depends on q, indicating the presence of multifractality. In other words, we want to point out how the two signals from completely different sources are cross-correlated in various time scales i.e. establishment of a direct correlation between the change in sound features to the change in EEG signal characteristics.

## RESULTS AND DISCUSSIONS:

For preliminary analysis, we chose five electrodes from the frontal and fronto-parietal lobe viz. F3, F4, Fp1, Fp2 and Fz, as the frontal lobe has been long associated with cognition of music and other higher order cognitive skills. The cross-correlation coefficient ($\gamma_x$) corresponding to the two experimental conditions were computed and the difference between the two conditions were computed and the corresponding graph (**Fig. 3**) shows the same. The complete 2 min signal (both EEG and audio signal) was segregated into 4 parts of 30 second each and for each part the cross-correlation coefficient was computed. **Fig. 3** represents the change in cross-correlation coefficient under the effect of drone stimulus. It is worth mentioning here that a decrease in the value of $\gamma_x$ signifies an increase in correlation between the two signals.

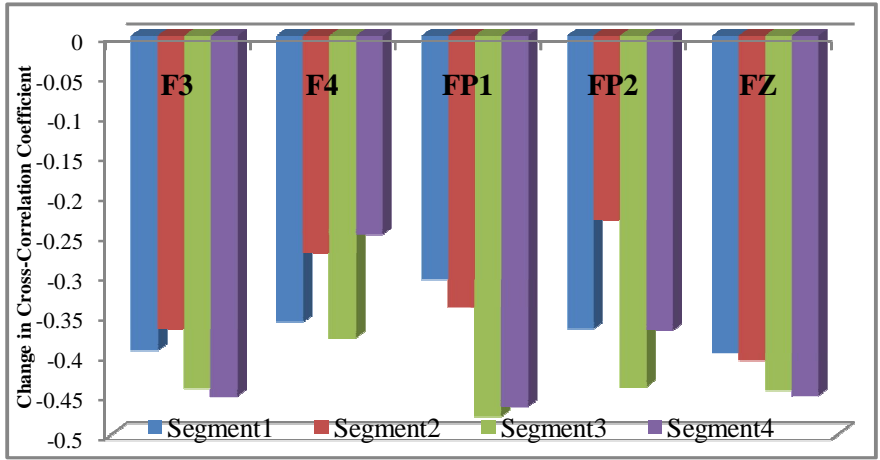

**Fig. 3: Variation of correlation between rest and music condition**

From the figure it is clear that the degree of correlation between audio signal and the EEG signal generated from the different electrodes increase with the progress of time. In most of the cases it is seen that the degree of correlation is the highest in the last third segment i.e. in between 1 min - 1 min 30 second and in few electrodes (i.e. Fz and F3), the degree of cross-correlation is highest in the last segment i.e. in between 1min 30 sec - 2 min. In our previous work [15], we reported how the complexity of the EEG signals generated from frontal lobes increase significantly under the influence of *tanpura* drone signal, while in this work we report for the first time how the audio signal which causes the change is directly correlated with the output EEG signal. Also, how the correlation varies during the course of the experiment is also an interesting observation from this experiment. From the figure, a gradual increase in the degree of cross-correlation is observed, but the 2nd part shows a fall in few electrodes, but in the 3rd and 4th part there is always an increase. It can thus be interpreted that the middle part of the audio signal is the most engaging part as it is in this section that the correlation between the frontal lobes and the audio signal becomes the highest. In our previous study also, we have shown that the increase in complexity was the highest in this part only. So the results from this unique experiment corroborates our previous findings. Next, we wanted to see how the inter/intra lobe cross-correlations vary under the effect of *tanpura* drone stimuli. So we calculated the cross-correlation coefficient between pairs of electrodes chosen for our study during the two experimental conditions i.e. "rest/no music state" and "with music" state. Again the difference between the two states have been plotted in **Fig. 4.** In this case we considered only the electrodes of left and right hemispheres and neglected the frontal midline electrode Fz.

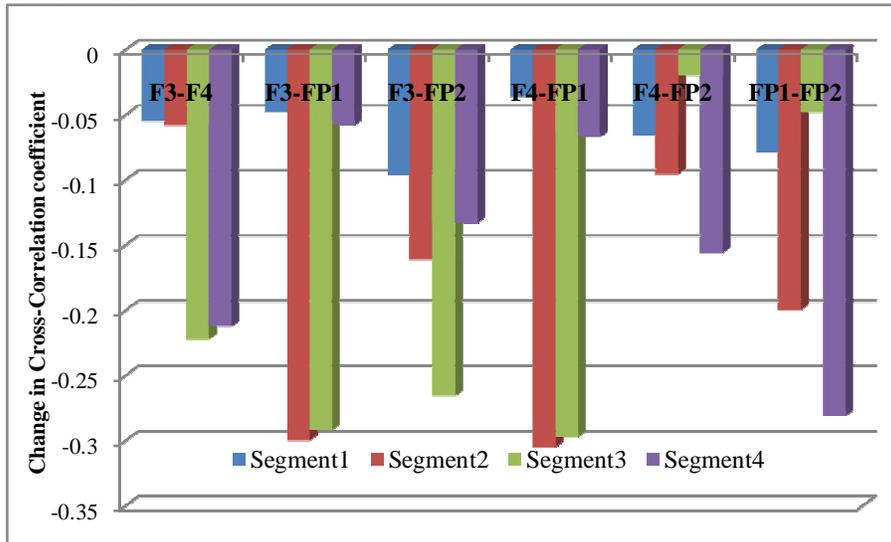

**Fig.4**: Variation of correlation among different electrodes

From the figure, it is seen that the correlation between left and right frontal electrode (i.e. F3 and F4) with the left fronto-parietal electrode i.e. FP1 has the most significant increase in the 2nd and 3rd segment of the audio stimuli. Apart from these, F3-FP2 correlation also increase consistently again in the 2nd and 3rd segments of the audio clip while inter lobe frontal correlation, i.e. between F3 and F4 electrodes show the highest rise in the last two parts of the audio signal. Inter-lobe fronto-parietal correlation rises significantly in the 2nd and 4th segments of the experiment. Thus, from the figure it is clear that different lobes of human brain activate themselves differently and in different portions under the effect of a simple auditory stimuli. In general, it is seen that the middle portion of a audio clip possesses the most important information which leads to higher degree of cross-correlation among the different lobes of the human brain. The last portion of the signal also engages a certain section of the brain to a great extent. This experiment provides novel insights into the complex neural dynamics going on in the human brain during the cognition and perception of an audio signal.

**CONCLUSION**
Professor Michael Ballam of Utah State University explains the effects of musical repetition: *"The human mind shuts down after three or four repetitions of a rhythm, or a melody, or a harmonic progression."* As a result, repetitive rhythmic music may cause people to actually release control of their thoughts, making them more receptive to whatever lyrical message is joined to the music. The *tanpura* drone in Hindustani music is a beautiful and most used example of repetitive music wherein the same pattern repeats itself again and again to engage the listeners and also to create an atmosphere. In this novel study, we deciphered a direct correlation between the source audio signal and the output EEG signal and also studied the correlation between different parts of the brain under the effect of same auditory stimulus. The following are the interesting conclusions obtained from the study:
1. For the first time, there is direct evidence of correlation existing between audio signal and the sonified (upsampled) EEG signals obtained from different lobes of human brain. The degree of correlation goes on increasing as the audio clip progresses and becomes maximum in the 3rd and 4th segment of the audio clip which is around 1-2 min in our case. The rise in correlation is different in scale in different electrodes, but in general we have found a stipulated time period wherein the effect of music on human brain is the maximum.

2. While computing the degree of correlation among different parts of the brain, we found that the audio clip has the ability to activate different brain regions simultaneously or in different times. Again, we find that the mid-portions of the audio clip are the ones which leads to most pronounced correlation in different electrode combinations. In the final portion of the audio clip also we find high value of $\gamma_x$ in several electrode combinations. This shows the ability of a music clip to engage several areas of the brain at a go not possible by any other stimulus at hand.

In conclusion, it can be said that this first-of-its kind study provides unique insights into the complex neural and audio dynamics simultaneously and has the potential to go a long way to device a methodology for scientific basis of cognitive music therapy. Future works going on in our Centre include the analysis of sonified EEG signals where emotional music have been used as a stimuli and development of a robust emotion classifier algorithm.

**ACKNOWLEDGEMENTS:**
One of the authors, AB acknowledges the Department of Science and Technology (DST), Govt. of India for providing (A.20020/11/97-IFD) the DST Inspire Fellowship to pursue this research work. SS acknowledges the Council of Scientific & Industrial Research (CSIR), Govt. of India for providing the Senior Research Fellowship to pursue this research (09/096(0876)/2017-EMR-I).